\documentclass[pdflatex,sn-mathphys-num]{sn-jnl}

\usepackage{graphicx}%
\usepackage{multirow}%
\usepackage{amsmath,amssymb,amsfonts}%
\usepackage{amsthm}%
\usepackage{mathrsfs}%
\usepackage[title]{appendix}%
\usepackage{xcolor}%
\usepackage{textcomp}%
\usepackage{manyfoot}%
\usepackage{booktabs}%
\usepackage{algorithm}%
\usepackage{algorithmicx}%
\usepackage{algpseudocode}%
\usepackage{listings}%
\usepackage{lscape}

\newcommand{\bbeta}{{\mbox{\boldmath $\beta$}}}

\newcommand{\bSigma}{{\mbox{\boldmath $\Sigma$}}}

\newcommand{\bgamma}{{\mbox{\boldmath $\gamma$}}}

\newcommand{\x}{{\mbox{\boldmath $x$}}}

\newcommand{\X}{{\mbox{\boldmath $X$}}}

\theoremstyle{thmstyleone}%
\theoremstyle{thmstyletwo}%

\theoremstyle{thmstylethree}%

\begin{document}

\title[Article Title]{Application of Propensity Score Models and Causal Estimators in Observational Studies under Model Misspecification}


\author*[1]{\fnm{Apu Chandra} \sur{Das}}\email{apdas@unmc.edu}

\author[2]{\fnm{Sakib} \sur{Salam}}\email{msalam@mcw.edu}

\author[3]{\fnm{Md Robiul Islam} \sur{Talukder}}\email{mtalukder@kumc.edu}

\author[4]{\fnm{Ashim Chandra} \sur{Das}}\email{ashim2305341488@diu.edu.bd}

\author[5]{\fnm{Antar Chandra} \sur{Das}}\email{antar2305341489@diu.edu.bd}

\author[6]{\fnm{Rakhi} \sur{Chowdhury}}\email{rachowdhury@unmc.edu}

\author[7]{for the Alzheimer’s Disease Neuroimaging Initiative}

\affil*[1]{\orgdiv{Department of Biostatistics}, \orgname{University of Nebraska Medical Center}, \orgaddress{\street{984375 Nebraska Medical Center}, \city{Omaha}, \postcode{68198}, \state{Nebraska}, \country{USA}}}

\affil[2]{\orgdiv{Division of Biostatistics}, \orgname{Medical College of Wisconsin}, \orgaddress{\street{8701 Watertown Plank Rd}, \city{Milwaukee}, \postcode{53226}, \state{Wisconsin}, \country{USA}}}

\affil[3]{\orgdiv{Department of Biostatistics and Data Science}, \orgname{University of Kansas Medical Center}, \orgaddress{\street{3901 Rainbow Boulevard}, \city{Kansas City}, \postcode{66160}, \state{Kansas}, \country{USA}}}

\affil[4,5]{\orgdiv{Department of Software Engineering}, \orgname{Daffodil International  University}, \orgaddress{\street{Daffodil Smart City}, \city{Savar}, \postcode{1216}, \state{Dhaka}, \country{Bangladesh}}}

\affil[6]{\orgdiv{Department of Pharmaceutical Sciences}, \orgname{University of Nebraska Medical Center}, \orgaddress{\street{984375 Nebraska Medical Center}, \city{Omaha}, \postcode{68198}, \state{Nebraska}, \country{USA}}}

\affil[7]{\orgdiv{Data used in preparation of this article were obtained from the Alzheimer’s Disease Neuroimaging Initiative (ADNI) database (adni.loni.usc.edu). As such, the investigators within the ADNI contributed to the design and implementation of ADNI and/or provided data but did not participate in analysis or writing of this report. A complete listing of ADNI investigators can be found at: http://adni.loni.usc.edu/wp-content/uploads/how\_to\_apply/ADNI\_Acknowledgement\_List.pdf}}


\abstract{Propensity score (PS) methods are widely used in observational studies to reduce confounding and estimate causal treatment effects. However, the validity of PS-based causal estimators depends heavily on correct model specification, and model misspecification may lead to substantial bias and instability. In this study, we systematically evaluate the performance of commonly used causal estimators, including response surface modeling (RSM), inverse probability weighting (IPW), and augmented inverse probability weighting (AIPW), under varying levels of PS and outcome model misspecification. We compare classical logistic regression with several machine learning approaches for PS estimation, including random forests (RF), support vector machines (SVM), and linear discriminant analysis (LDA). Extensive simulation studies were conducted under multiple scenarios defined by combinations of correctly specified and misspecified PS and outcome models, varying sample sizes, and different covariate correlation structures. Estimator performance was assessed using bias, absolute bias, root mean squared error, empirical standard error, and confidence interval width. Results demonstrate that AIPW consistently provides robust and stable estimates across most scenarios due to its doubly robust property, whereas IPW is highly sensitive to PS misspecification and unstable PS estimates produced by flexible machine learning methods. RSM performs well only when the outcome model is correctly specified. Real-world applications using the ACTG175 clinical trial and the Alzheimer’s Disease Neuroimaging Initiative (ADNI) dataset further illustrate the practical implications of estimator choice and PS modeling strategy. Overall, our findings highlight the importance of integrating flexible machine learning approaches within doubly robust frameworks to improve causal effect estimation in observational studies.}


\maketitle

\section{Introduction}\label{intro}

While randomized controlled trials (RCTs) are considered the gold standard for assessing treatment efficacy, ethical, financial, and logistical constraints often limit their feasibility in biomedical and clinical research \cite{hernan2004definition, imbens2015causal}. As a result, observational studies and real-world data have become increasingly important for evaluating treatment effects in practice. The assignment of treatment in observational studies is not random, which leads to confounding. Patients receiving different treatments may differ systematically in their baseline characteristics, which can bias treatment effect estimates if not properly addressed. The potential outcomes framework provides a formal foundation for addressing these challenges by guiding adjustment for such imbalances \cite{rubin1974estimating, splawa1990application}.

Over the past decades, a wide variety of statistical methods have been developed to estimate causal effects under these conditions. Early foundational work focused on standardization and outcome regression approaches, which were formalized by Cochran \cite{cochran1968}. The paradigm shifted significantly with the introduction of the propensity score (PS) framework by Rosenbaum and Rubin \cite{rosenbaum1983}, establishing a principled method to control for confounding through balancing scores. This milestone catalyzed the development of PS matching, subclassification, and stratification methods further advanced by Imbens \cite{imbens2004} and others \cite{abadie2011,hansen2004}. To address time-varying confounding, Robins et al. \cite{robins2000marginal} subsequently introduced inverse probability weighting (IPW) within the marginal structural model framework. This line of semiparametric research, rooted in earlier work by Robins et al. \cite{robins1994estimation}, ultimately culminated in doubly robust architectures like the augmented inverse probability weighting (AIPW) estimator formalized by Bang and Robins \cite{bang2005doubly}. 

More recently, van der Laan et al. \cite{vanderlaan2011} proposed targeted maximum likelihood estimation (TMLE), offering a flexible semiparametric framework that explicitly integrates machine learning algorithms with formal causal inference.

Commonly used methods in causal inference include response surface model (RSM), inverse probability weighting (IPW), and augmented inverse probability weighting (AIPW). RSM models the relationship between covariates and outcomes directly, while IPW relies on PS—the probability of receiving treatment given observed covariates—to construct a weighted pseudo-population in which treatment assignment is independent of covariates \cite{rosenbaum1983}. AIPW combines both outcome modeling and PS weighting and is particularly attractive due to its doubly robust property, providing consistent estimates if either the outcome model or the PS model is correctly specified \cite{bang2005doubly}.
The PS is therefore a central tool in causal inference to deal with the confounding issue through covariate balancing \cite{rosenbaum1983}. Its application has recently extended to areas such as real-world data integration and information borrowing. In particular, Wang et al. \cite{wang2019} and Lu et al. \cite{lu2022} developed PS–integrated power prior approaches to incorporate external data into clinical studies, while Das et al. \cite{das2026a, das2026b} proposed flexible borrowing frameworks to improve operating characteristics in clinical trial settings. These developments highlight the growing importance of PS methodology in modern clinical research. However, the performance of PS-based approaches critically depends on accurate estimation of the PSs. In particular, misspecification of the PS model can lead to inadequate covariate balance, resulting in biased estimates and increased variability \cite{austin2011, kang2007demystifying}. Achieving accurate estimation typically requires correct specification of the functional form of the treatment assignment mechanism, including appropriate handling of nonlinearities and interactions. Traditionally, PSs are estimated using parametric models such as logistic regression. While these models are simple and interpretable, they may fail to capture complex relationships between covariates and treatment assignment, potentially leading to model misspecification \cite{lee2010improving}. To address this limitation, machine learning methods such as random forests (RF), support vector machines (SVM), and linear discriminant analysis (LDA) have been increasingly adopted for PS estimation \cite{mccaffrey2013}. Although these approaches can improve predictive performance, they may also yield extreme PS estimates, leading to unstable weights and increased variability in IPW estimators \cite{lee2010improving, austin2011}.

Despite these developments, misspecification remains a fundamental challenge in observational studies, as the true functional forms of both the outcome and PS models are typically unknown. It may arise from incorrect functional forms, omitted variables, or unaccounted interactions, potentially leading to biased estimates and increased variability. Several studies have demonstrated that causal estimators are sensitive to model misspecification, with RSM and IPW performing poorly under incorrect specification and doubly robust methods offering only partial protection \cite{bang2005doubly, kang2007demystifying, waernbaum2012, hainmueller2012, kurz2022}. For instance, Kang and Schafer \cite{kang2007demystifying} showed that even doubly robust estimators can exhibit substantial bias and instability in practical settings. Similarly, Waernbaum \cite{waernbaum2012} demonstrated that no single estimator uniformly dominates across misspecification scenarios, and Hainmueller \cite{hainmueller2012} highlighted limitations of traditional PS methods and proposed alternative weighting strategies. However, the impact of machine learning–based PS estimation on the performance of classical causal estimators under model misspecification remains insufficiently understood. Recent work by Chen et al. \cite{chen2025} compared several causal estimators combined with super learner approaches and demonstrated improved performance of doubly robust methods under certain settings. Building on this line of work, a more systematic evaluation across a broader range of model misspecification scenarios, sample sizes, and covariate correlation structures is needed. Therefore, our primary objective is to systematically evaluate the impact of machine learning–based PS estimation methods on the performance of commonly used causal estimators—including RSM, IPW, and AIPW—under varying levels of model specification, sample sizes, and covariate correlation structures.

In this study, we design extensive simulation scenarios to evaluate the performance of causal estimators under varying model specifications and data settings. We consider four model specification regimes based on combinations of correctly specified and misspecified PS and outcome models. 
Across these misspecification settings, we compare how machine learning–based PS estimation strategies influence the performance of causal estimators in terms of bias, variability, and robustness. We further illustrate these comparisons using two real-world datasets: the ACTG175 and ADNI studies.

The remainder of the paper is organized as follows. Section \ref{causal_framework} presents the causal inference framework and estimation procedures. Section \ref{simulation} describes simulation design, including scenario construction, data-generating mechanisms, and simulation results. Section \ref{application} presents the application to real data. Finally, Section \ref{discussion} concludes with a discussion of key findings and implications for practice.

\section{Causal Inference Framework and Estimation} \label{causal_framework}

We adopt the potential outcomes framework for causal inference, originally developed by Neyman \cite{splawa1990application} and Rubin \cite{rubin1974estimating}. Consider a sample of $n$ independent subjects by $i=1,\dots, n$. Let $A_i \in \{0,1\}$ denote a binary treatment indicator, where $A_i=1$  corresponds to exposure to the treatment and $A_i=0$ denotes control. Let $\X_i$ be a set of observed covariates and $Y_i$ denote a continuous outcome of interest. For each subject $i$, define the potential outcomes $Y_i(1) \ \text{and} \ Y_i(0)$ representing the outcomes that would be observed under treatment and control, respectively. 

\begin{equation*}
    Y_i(A_i) = 
    \begin{cases} 
    Y_i(1) & \text{if } A_i=1, \\
    Y_i(0) & \text{if } A_i=0.
    \end{cases}
\end{equation*}

The observed outcome is related to the potential outcomes through
\begin{equation*}
    Y_i=A_iY_i(1)+(1-A_i)Y_i(0).
\end{equation*}

Because only one of the two potential outcomes is observed for each subject, causal inference is fundamentally a missing data problem, often referred to as the fundamental problem of causal inference \cite{holland1986statistics, gelman2011causality}. The individual-level causal effect is defined as $Y_i(1)-Y_i(0)$. The distribution of $Y_i(A)$ characterizes the outcomes that would be observed in the population if all individuals were assigned to treatment $(A=1)$ or control $(A=0)$, with corresponding means $\mathbb{E}[Y(1)]$ and $\mathbb{E}[Y(0)]$. Because estimating causal effects at the individual level is typically infeasible, our focus is on well-defined aggregate causal effects. While individual causal effects are expressed as differences between potential outcomes, aggregated causal effects are generally defined as averages of these individual-level differences. The most general of these is the average treatment effect (ATE) for the entire population,

\begin{equation*}
    \tau_{ATE}=\mathbb{E} \bigl[Y(1)-Y(0)\bigr].
\end{equation*}

Identification of ATE relies on three standard assumptions. First, the stable unit treatment value assumption (SUTVA) \cite{angrist1996identification} holds, implying that each subject’s potential outcome depends only on its own treatment assignment and that there are no hidden versions of the treatment. Careful study design can minimize violations, e.g., by clearly defining the treatment, ruling out interference between subjects, and restricting the population to one with a single treatment version \cite{imbens2015causal}. Second, we assume conditional exchangeability (also referred to as unconfoundedness), such that conditional on the observed covariates $\X$, the treatment assignment is independent of the potential outcomes, 
\begin{equation*}
    \bigl\{Y(1), Y(0)\bigr\} \perp A \mid \X. 
\end{equation*}
This assumption states that the unobserved potential outcome corresponding to the treatment level not received by an individual can be inferred from outcomes observed among individuals with comparable measured covariates. Equivalently, under exchangeability, treatment assignment provides no additional information about the potential outcomes beyond what is explained by the observed covariates $\X$, implying that adjustment for $\X$ removes confounding. Exchangeability requires that treatment choices are independent of the outcomes that would occur under alternative treatment conditions. Consequently, when estimating a treatment effect, any observed difference in outcomes between treated and untreated groups must be attributable to the treatment itself rather than to preexisting group differences unrelated to the intervention \cite{rubin1974estimating, hernan2004definition}. Third, the positivity (overlap) assumption requires that every subject has a nonzero probability of receiving each treatment level given its covariates, that is, 
\begin{equation*}
    0< \Pr(A=1 \mid \X) <1
\end{equation*}
almost surely. Together, these assumptions ensure that counterfactual outcomes are well defined, identifiable, and estimable from the observed data via

\begin{equation}
    \tau_{\mathrm{ATE}} = \mathbb{E}\Big[ \mathbb{E}(Y \mid A=1, \X) - \mathbb{E}(Y \mid A=0, \X) \Big].
\end{equation}

\subsection{Estimating the treatment effect} \label{est_treat_effect}
One common approach to estimating the ATE is to model the conditional mean outcome, also known as the response surface model (RSM) \cite{austin2011tutorial}. Any regression models, including machine learning or more advanced models, can be used to fit RSM \cite{athey2016recursive, hill2011bayesian}. Assuming $\mu(a, \x) = \mathbb{E}(Y \mid A = a, \X = \x), \quad a \in \{0,1\}$, we get the formula for the treatment effect:
\begin{equation*}
    \tau_{\mathrm{ATE}}=\mathbb{E} \Big[\mu(1, \X) - \mu(0,\X) \Big]
\end{equation*}
If consistent estimators $\hat{\mu}(1,\x)$ and $\hat{\mu}(0,\x)$ are available, the ATE can be estimated as

\begin{equation}
    \widehat{\tau}_{\mathrm{RSM}} = \frac{1}{n} \sum_{i=1}^n \Big\{ \hat{\mu}(1, \X_i) - \hat{\mu}(0, \X_i) \Big\}.
\end{equation}

This estimator relies critically on the correct specification of the outcome model. In practice, misspecification of the regression function—especially under limited covariate overlap—can lead to biased estimates and understated uncertainty \cite{king2006dangers}. These concerns motivate alternative estimators that rely on modeling the treatment assignment mechanism rather than the outcome process \cite{glynn2010introduction}.

\subsection{Inverse probability weighting} \label{ipw}
Inverse probability weighting (IPW) addresses confounding by reweighting observations based on the inverse probability of receiving the observed treatment \cite{hirano2003efficient}. Define the propensity score $e(\X) = \Pr(A = 1 \mid \X)$ that represents the conditional probability of receiving the treatment given the observed covariates. Under the identification assumptions, the ATE can be written as

\begin{equation*}
    \tau_{\mathrm{ATE}} = \mathbb{E}\left[ \frac{A Y}{e(\X)} - \frac{(1-A) Y}{1 - e(\X)} \right].
\end{equation*}
Replacing the unknown PS with an estimate $\hat{e}(\X)$, the IPW estimator is given by

\begin{equation}
    \widehat{\tau}_{\mathrm{IPW}} = \frac{1}{n} \sum_{i=1}^n \left( \frac{A_i Y_i}{\hat{e}(\X_i)} - \frac{(1-A_i) Y_i}{1 - \hat{e}(\X_i)} \right) \label{eqn:ipw}.
\end{equation}
In equation \eqref{eqn:ipw}, subjects in the treatment group receive a weight of $1/\hat{e}(\X_i)$ and those in the control group receive a weight of $1/(1-\hat{e}(\X_i))$. Intuitively, IPW constructs a pseudo-population in which treatment assignment is independent of measured covariates. Theoretically, if the PS is known, then the IPW estimator is unbiased. In that case, $e(\X)$ must be the true PS for this estimator to be consistent \cite{tsiatis2006semiparametric}. While consistent when the PS model is correctly specified, IPW estimators can suffer from high variance, particularly when estimated PSs are close to 0 or 1, resulting in extreme weights.

\subsection{Augmented inverse probability weighting} \label{aipw}
The augmented inverse probability weighting (AIPW) estimator combines outcome regression and inverse probability weighting into a single estimating equation \cite{robins1995analysis}. Let $\hat{\mu}(a, \X)$ be estimators of $\mu(a, \X)$ and $\hat{e}(\X)$ an estimator of the PS.

The AIPW estimating function for subject $i$ is

\begin{multline}
            \hat \psi_i^{\mathrm{AIPW}} = \left[ \frac{A_i Y_i}{\hat{e}(\X_i)} - \frac{(1-A_i) Y_i}{1 - \hat{e}(\X_i)} \right] - \left[ \frac{A_i - \hat{e}(\X_i)}{\hat{e}(\X_i)(1-\hat{e}(\X_i))} \right] \cdot \\ \Big[ (1-\hat{e}(\X_i)) \hat{\mu}(1,\X_i) + \hat{e}(\X_i) \hat{\mu}(0,\X_i) \Big]
\end{multline}
The AIPW estimator of the ATE is then

\begin{equation*}
    \widehat{\tau}_{\mathrm{AIPW}} = \frac{1}{n} \sum_{i=1}^n \hat \psi_i^{\mathrm{AIPW}}.
\end{equation*}
A key property of this estimator is double robustness: $\widehat{\tau}_{\mathrm{AIPW}} \xrightarrow{p} \tau_{\mathrm{ATE}} $ if either $\hat{e}(\X)$ or $\hat{\mu}(a,\X)$ is correctly specified (the correctly specified model converges at a parametric rate \cite{bang2005doubly}).

That is, consistency is guaranteed provided at least one of the two nuisance models is correctly specified, though not necessarily both. When both models are correct, the AIPW estimator is asymptotically efficient within the class of regular estimators. The AIPW estimator can be viewed as an IPW estimator with a bias-correcting augmentation term that stabilizes estimation and reduces variance. Asymptotic variance can be estimated using sandwich estimators or nonparametric bootstrap procedures.

\section{Simulation} \label{simulation}
\subsection{Objectives}
The objective of this simulation study is to evaluate the performance of several causal effect estimators under varying levels of model specification and data complexity. Specifically, we aim to compare IPW, AIPW, and RSM when PS is estimated using different approaches, including classical logistic regression and machine learning methods. By systematically introducing misspecification in the PS and outcome models, and by varying sample size and covariate correlation, we seek to quantify the impact of misspecification on the bias, variance, and robustness of these estimators of the true ATE.

\subsection{Data generation and simulation mechanism}
We conduct an extensive simulation study to evaluate the finite-sample performance of several causal estimators under different levels of model misspecification. For each replicate, we generate independent observations $(Y_i, A_i, \X_i)$, where $A_i \in \{0,1\}$ denotes treatment, $Y_i$ is a continuous outcome, and $\X_i = (X_{i1},\ldots,X_{ip})^\top$ is a vector of baseline covariates, for $i = 1,\ldots,n$, with covariate dimension $p=9$. Covariates are generated from a multivariate normal distribution, $\X_i \sim \mathcal{N}_p(\mathbf{0}, \bSigma), \hspace{2mm} \Sigma_{jk} = \rho^{|j-k|}$, where $\rho \in \{0.2, 0.7\}$ controls the correlation among covariates. The treatment assignment follows a logistic regression model, $\Pr(A_i = 1 \mid \X_i) = \text{logit}^{-1}\!\left(\beta_0 + \sum_{j=1}^{p} \beta_j X_{ij}\right)$, with $\beta_0 = 0$ and fixed regression coefficients $\bbeta = (1.2, -0.8, 0.6, 1.1, -0.9, 0.5, 0.7, -1.0, -0.9)^\top$. Then, $Y_i$ is generated according to

\begin{equation*}
    Y_i = \alpha_0 + \tau A_i + \sum_{j=1}^{p} \gamma_j X_{ij} + \varepsilon_i, \hspace{3mm} \varepsilon_i \sim \mathcal{N}(0,1),
\end{equation*}
where $\alpha_0=0$, the true average treatment effect is set to $\tau = 2$ and $\bgamma = (0.6, -0.4, 1.1, 0.5, -1.2, 0.9, -0.3, -0.8, 1.5)^\top$.

We estimate PS using logistic regression (LR), random forests (RF), linear discriminant analysis (LDA), and support vector machines (SVM). PS misspecification is introduced by excluding two true confounders, $X_8$ and $X_9$, from the fitted models and by augmenting the linear predictor with interaction terms,
\begin{multline*}
    e(\X_i) = \Pr(A_i = 1 \mid \X_i) = \text{logit}^{-1}\!\left(
\eta_0 + \sum_{j \in \mathcal{S}_{\text{PS}}} \eta_j X_{ij}
+ \eta_{12} X_{i1}X_{i2}
+ \eta_{13} X_{i1}X_{i3}
\right), \\ \mathcal{S}_{\text{PS}}=\{1,\ldots,7\}.
\end{multline*}

This structure reflects both omitted-variable bias and incorrect functional form in the PS model. We replace PS values less than 0.025 and greater than 0.975 with 0.025 and 0.975, respectively, to avoid extreme weights. For outcome modeling, a linear regression model is fitted to estimate $\mathbb{E}(Y \mid A, \X)$. Outcome model misspecification is induced by omitting two covariates, $X_6$ and $X_7$, from the fitted model, yielding
\begin{equation*}
\mathbb{E}(Y_i \mid A_i, \X_i) = \delta_0 + \tau A_i + \sum_{j \in \mathcal{S}_{\text{out}}} \delta_j X_{ij}, \hspace{3mm}  \mathcal{S}_{\text{out}}=\{1,\ldots,5,8,9\}.
\end{equation*}

Predicted potential outcomes $\hat \mu(1, \X_i)$ and $\hat \mu(0, \X_i)$ are obtained by setting $A_i=1$ and $A_i=0$, respectively, and are used for both RSM estimation and as inputs to the AIPW estimator. We consider four model specification regimes: (i) both PS and outcome models correctly specified; (ii) misspecified PS with correctly specified outcome model; (iii) correctly specified PS with misspecified outcome model; and (iv) both models misspecified. In addition, we examined four data settings defined by sample size and covariate correlation, with $n \in \{200, 1000\}$ and $\rho \in \{0.2, 0.7\}$. In total, we test 16 different scenarios.

\subsection{Evaluation metrics}

For each simulation setting, we generated $B = 1000$ Monte Carlo replicates. Estimator performance was evaluated using several commonly used metrics for treatment effect estimation. Let $\hat{\tau}_b$ denote the estimated treatment effect from the $b^{th}$ replicate and let $\tau$ denote the true treatment effect. We computed the following quantities across replicates: bias, defined as (i)
$\text{Bias} = \frac{1}{B}\sum_{b=1}^{B} (\hat{\tau}_b - \tau)$, $b=1, \dots, B$; absolute bias, defined as (ii)
$\text{ABias} = \frac{1}{B}\sum_{b=1}^{B} \left|\hat{\tau}_b - \tau\right|$;
root mean squared error, defined as (iii)
$\text{RMSE} = \sqrt{\frac{1}{B} \sum_{b=1}^{B} (\hat{\tau}_b-\tau)^2}$;
the empirical standard error, defined as (iv)
$\text{SE} = \sqrt{\frac{1}{B-1} \sum_{b=1}^{B} (\hat{\tau}_b - \frac{1}{B}\sum_{\ell=1}^{B} \hat{\tau}_\ell)^2}$; and (v) the average width of 95\% confidence intervals (Width).

\subsection{Results}
\begin{figure} [ht]
\centering
\includegraphics[width=13cm,height=15cm]{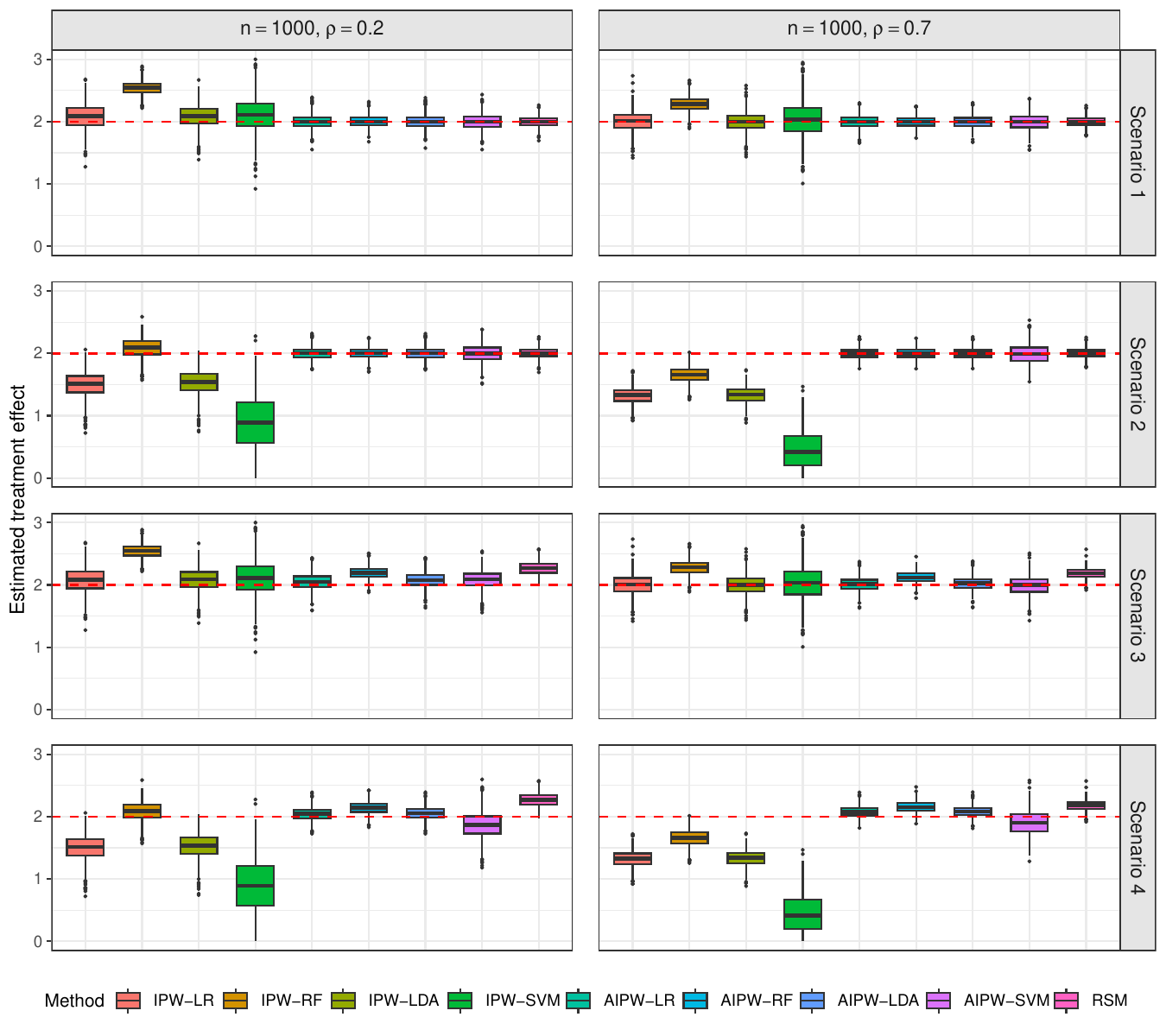}%
\caption{Boxplots of treatment effect estimates across 1,000 Monte Carlo replicates for $n=1000$. The dashed horizontal line indicates the true treatment effect $\tau=2$}%
\label{fig:boxplot-n1000}%
\end{figure}

Tables \ref{tab:sim_results_n200_rho0.2} and \ref{tab:sim_results_n200_rho0.7} summarize the simulation results for $n = 200$ with $\rho = 0.2$ and $\rho = 0.7$, respectively, while the corresponding results for $n = 1000$ are reported in Tables 1 and 2 of the Supplementary Materials. Figures \ref{fig:boxplot-n200} and \ref{fig:boxplot-n1000} display boxplots of the treatment effect estimates across Monte Carlo replicates for $n=200$ and $n=1000$, respectively. Across all scenarios, estimator performance is evaluated using bias, absolute bias, RMSE, SE, and the average width of the 95\% confidence interval.

\begin{figure} [ht]
\centering
\includegraphics[width=13cm,height=15cm]{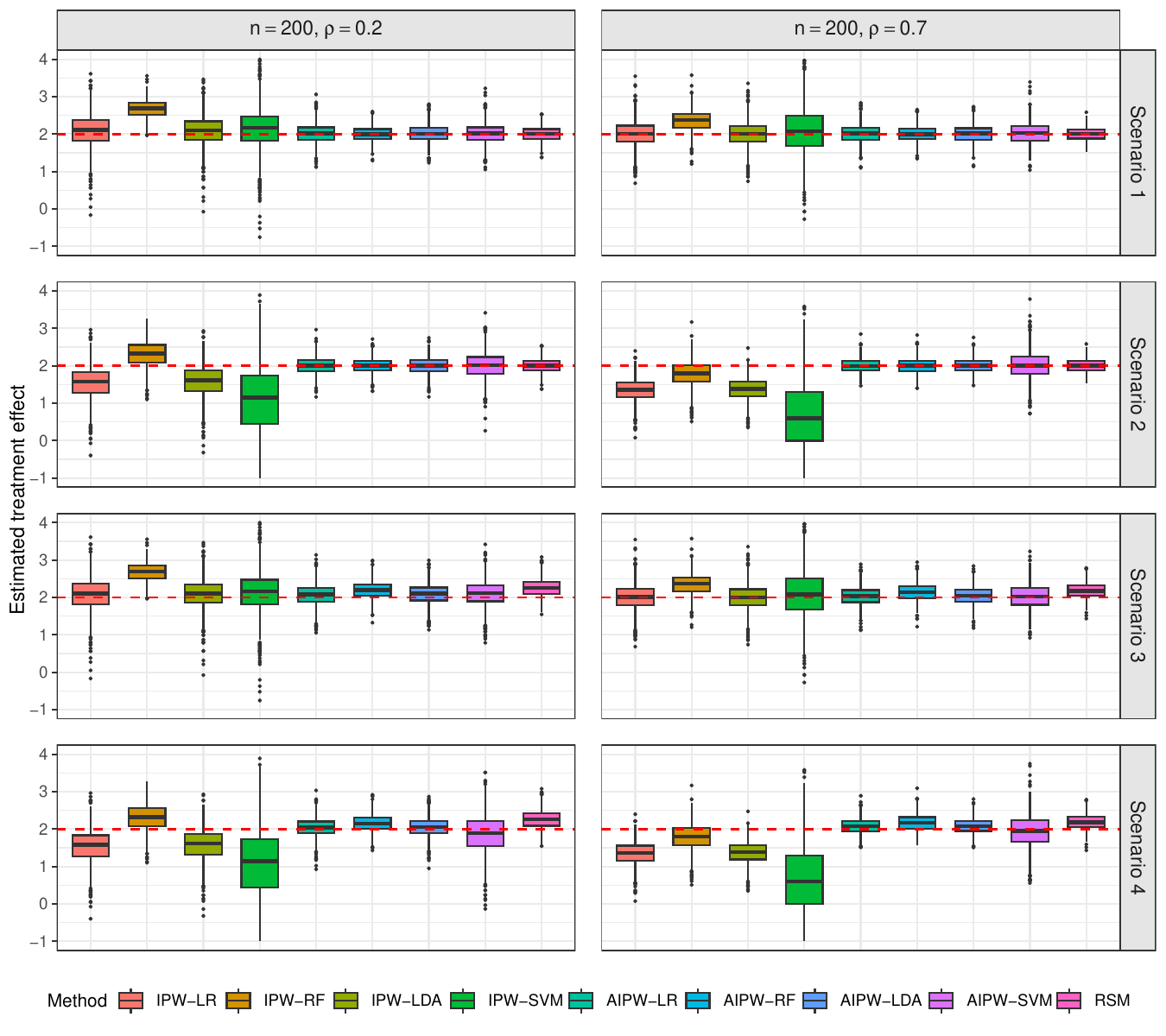}%
\caption{Boxplots of treatment effect estimates across 1,000 Monte Carlo replicates for $n=200$. The dashed horizontal line indicates the true treatment effect $\tau=2$}%
\label{fig:boxplot-n200}%
\end{figure}

When both the PS and outcome models are correctly specified (Scenario 1), AIPW and RSM consistently produce nearly unbiased estimates of the treatment effect, with small RMSE, SE, and narrow confidence intervals across all PS estimation methods. In contrast, IPW exhibits noticeable bias when the PS is estimated using more flexible machine learning approaches, particularly RF $(|Bias| \approx 0.68)$, with more modest but still elevated bias for SVM $(|Bias| \approx 0.16)$. This behavior is evident in both the tables and the boxplots, where IPW estimates display increased variability and systematic deviation from the true effect. These findings reflect the well-known sensitivity of IPW to PS estimation, especially when machine learning methods yield estimated scores close to the boundaries, resulting in unstable weights \cite{kang2007demystifying, robins2000marginal}.

\begin{table}[ht]
\centering
\caption{Simulation results across scenarios, PS, and estimators when $n=200$ and $\rho=0.2$}
\label{tab:sim_results_n200_rho0.2}
\begin{tabular}{c c c c c c c c c }
\toprule
Scenario &
PS model &
Estimator &
Mean &
Bias &
ABias &
RMSE &
SE &
Width \\
\midrule

\multirow{9}{*}{1}
& LR & IPW  & 2.083 & $-0.083$ & 0.365 & 0.475 & 0.468 &  1.823\\
& RF & IPW  & 2.684 & $-0.684$ & 0.684 & 0.723 & 0.235 &  0.914\\
& LDA & IPW & 2.089 & $-0.089$ & 0.325 & 0.428 & 0.419 &  1.710\\
& SVM & IPW & 2.162 & $-0.162$ & 0.491 & 0.663 & 0.643 &  2.662\\
\cmidrule(lr){2-9}
& LR & AIPW & 2.017 & $-0.017$ & 0.202 & 0.259 & 0.259 & 0.997\\
& RF & AIPW & 2.002 & $-0.002$ & 0.160 & 0.201 & 0.201 &  0.811\\
& LDA & AIPW & 2.013 & $-0.013$ & 0.188 & 0.240 & 0.240 &  0.936\\
& SVM & AIPW & 2.016 & $-0.016$ & 0.212 & 0.275 & 0.274 &  1.109\\
\cmidrule(lr){2-9}
& --  & RSM  & 2.002 & $-0.002$ & 0.151 & 0.190 & 0.190 & 0.767\\

\midrule
\multirow{9}{*}{2}
& LR & IPW  & 1.537 & 0.463 & 0.521 & 0.650 & 0.456 & 1.804\\
& RF & IPW  & 2.313 & $-0.313$ & 0.384 & 0.468 & 0.348 & 1.381\\
& LDA & IPW & 1.576 & 0.424 & 0.483 & 0.609 & 0.437 & 1.689\\
& SVM & IPW & 0.964 & 1.036 & 1.202 & 1.514 & 1.105 & 4.517\\
\cmidrule(lr){2-9}
& LR & AIPW & 2.003 & $-0.003$ & 0.176 & 0.224 & 0.224 & 0.897\\
& RF & AIPW & 2.001 & $-0.001$ & 0.159 & 0.203 & 0.203 & 0.825\\
& LDA & AIPW & 2.002 & $-0.002$ & 0.171 & 0.218 & 0.218 & 0.880\\
& SVM & AIPW & 2.006 & $-0.006$ & 0.277 & 0.354 & 0.354 & 1.380\\
\cmidrule(lr){2-9}
& -- & RSM  & 2.002 & $-0.002$ & 0.151 & 0.190 & 0.190 & 0.767\\

\midrule
\multirow{9}{*}{3}
& LR  & IPW & 2.083 & $-0.083$ & 0.365 & 0.475 & 0.468 & 1.523\\
& RF & IPW  & 2.684 & $-0.684$ & 0.684 & 0.723 & 0.235 & 0.914\\
& LDA & IPW  & 2.089 & $-0.089$ & 0.325 & 0.428 & 0.419 & 1.710\\
& SVM & IPW  & 2.162 & $-0.162$ & 0.490 & 0.663 & 0.643 & 2.662\\
\cmidrule(lr){2-9}
& LR & AIPW & 2.070 & $-0.069$ & 0.240 & 0.306 & 0.298 & 1.219\\
& RF & AIPW & 2.195 & $-0.195$ & 0.250 & 0.305 & 0.235 & 0.926\\
& LDA & AIPW & 2.092 & $-0.092$ & 0.231 & 0.291 & 0.277 & 1.113\\
& SVM & AIPW & 2.091 & $-0.091$ & 0.286 & 0.365 & 0.354 & 1.458\\
\cmidrule(lr){2-9}
& -- & RSM  & 2.259 & $-0.259$ & 0.293 & 0.356 & 0.245 & 0.953\\

\midrule
\multirow{9}{*}{4}
& LR & IPW  & 1.537 & 0.463 & 0.521 & 0.650 & 0.456 & 1.804\\
& RF & IPW  & 2.313 & $-0.313$ & 0.384 & 0.468 & 0.348 & 1.381\\
& LDA & IPW & 1.576 & 0.424 & 0.483 & 0.609 & 0.437 & 1.689\\
& SVM & IPW & 0.964 & 1.036 & 1.202 & 1.514 & 1.105 & 4.517\\
\cmidrule(lr){2-9}
& LR & AIPW & 2.047 & $-0.047$ & 0.207 & 0.267 & 0.263 & 1.067\\
& RF & AIPW & 2.155 & $-0.155$ & 0.220 & 0.276 & 0.228 & 0.907\\
& LDA & AIPW & 2.055 & $-0.055$ & 0.203 & 0.261 & 0.255 & 1.027\\
& SVM & AIPW & 1.868 & $0.132$ & 0.418 & 0.545 & 0.529 & 2.075\\
\cmidrule(lr){2-9}
& -- & RSM  & 2.259 & $-0.259$ & 0.293 & 0.356 & 0.245 & 0.953\\

\bottomrule
\end{tabular}
\end{table}

\begin{table}[ht]
\centering
\caption{Simulation results across scenarios, PS, and estimators when $n=200$ and $\rho=0.7$}
\label{tab:sim_results_n200_rho0.7}
\begin{tabular}{c c c c c c c c c }
\toprule
Scenario &
PS model &
Estimator &
Mean &
Bias &
ABias &
RMSE &
SE &
Width \\
\midrule

\multirow{9}{*}{1}
& LR & IPW  & 2.007 & $-0.007$ & 0.278 & 0.361 & 0.362 & 1.456\\
& RF & IPW  & 2.359 & $-0.359$ & 0.386 & 0.455 & 0.280 & 1.123\\
& LDA & IPW & 1.999 & $0.001$ & 0.262 & 0.340 & 0.340 & 1.373\\
& SVM & IPW & 2.131 & $-0.131$ & 0.546 & 0.743 & 0.732 & 3.014\\
\cmidrule(lr){2-9}
& LR & AIPW & 2.012 & $-0.012$ & 0.188 & 0.237 & 0.237 & 0.902\\
& RF & AIPW & 2.002 & $-0.002$ & 0.168 & 0.210 & 0.210 & 0.796\\
& LDA & AIPW & 2.011 & $-0.011$ & 0.179 & 0.225 & 0.225 & 0.871\\
& SVM& AIPW & 2.023 & $-0.023$ & 0.234 & 0.299 & 0.298 & 1.184\\
\cmidrule(lr){2-9}
& -- & RSM  & 2.004 & $-0.004$ & 0.145 & 0.180 & 0.180 & 0.685\\

\midrule
\multirow{9}{*}{2}
& LR & IPW  & 1.357 & 0.643 & 0.647 & 0.711 & 0.303 & 1.179\\
& RF & IPW  & 1.787 & 0.213 & 0.318 & 0.409 & 0.349 & 1.431\\
& LDA & IPW  & 1.377 & 0.623 & 0.626 & 0.688 & 0.293 & 1.141\\
& SVM & IPW  & 0.388 & 1.612 & 1.668 & 1.979 & 1.148 & 4.559\\
\cmidrule(lr){2-9}
& LR & AIPW & 2.006 & $-0.006$ & 0.156 & 0.195 & 0.195 & 0.750\\
& RF & AIPW & 2.001 & $-0.001$ & 0.164 & 0.205 & 0.205 & 0.779\\
& LDA & AIPW & 2.006 & $-0.006$ & 0.154 & 0.192 & 0.192 & 0.749\\
& SVM& AIPW & 2.018 & $-0.018$ & 0.289 & 0.373 & 0.373 & 1.469\\
\cmidrule(lr){2-9}
& -- & RSM & 2.004 & $-0.004$ & 0.145 & 0.180 & 0.180 & 0.685\\

\midrule
\multirow{9}{*}{3}
& LR & IPW  & 2.007 & $-0.007$ & 0.278 & 0.361 & 0.362 & 1.456\\
& RF & IPW  & 2.359 & $-0.359$ & 0.386 & 0.455 & 0.280 & 1.123\\
& LDA & IPW  & 1.999 & 0.001 & 0.262 & 0.340 & 0.340 & 1.373\\
& SVM & IPW  & 2.131 & $-0.131$ & 0.546 & 0.743 & 0.732 & 3.014\\
\cmidrule(lr){2-9}
& LR & AIPW & 2.034 & $-0.03$ & 0.197 & 0.249 & 0.247 & 0.948\\
& RF & AIPW & 2.138 & $0.138$ & 0.213 & 0.267 & 0.229 & 0.902\\
& LDA & AIPW & 2.045 & $-0.045$ & 0.191 & 0.239 & 0.235 & 0.908\\
& SVM& AIPW & 2.017 & $-0.017$ & 0.264 & 0.336 & 0.336 & 1.346\\
\cmidrule(lr){2-9}
& --  & RSM  & 2.184 & $-0.184$ & 0.227 & 0.277 & 0.207 & 0.815\\

\midrule
\multirow{9}{*}{4}
& LR & IPW  & 1.357 & 0.643 & 0.647 & 0.711 & 0.303 & 1.179\\
& RF & IPW  & 1.787 & 0.213 & 0.318 & 0.409 & 0.349 & 1.431\\
& LDA & IPW & 1.377 & 0.623 & 0.626 & 0.688 & 0.293 & 1.141\\
& SVM & IPW  & 0.388 & 1.612 & 1.668 & 1.979 & 1.148 & 4.559\\
\cmidrule(lr){2-9}
& LR & AIPW & 2.083 & $-0.083$ & 0.173 & 0.219 & 0.203 & 0.812\\
& RF & AIPW & 2.164 & $-0.164$ & 0.226 & 0.278 & 0.225 & 0.868\\
& LDA & AIPW & 2.086 & $-0.086$ & 0.171 & 0.217 & 0.199 & 0.785\\
& SVM& AIPW & 1.948 & 0.052 & 0.360 & 0.458 & 0.455 & 1.866\\
\cmidrule(lr){2-9}
& -- & RSM  & 2.184 & $-0.184$ & 0.227 & 0.277 & 0.207 & 0.815 \\

\bottomrule
\end{tabular}
\end{table}

Under PS misspecification with a correctly specified outcome model (Scenario 2), AIPW and RSM continue to demonstrate strong performance, yielding estimates with negligible bias and favorable RMSE and confidence interval width. In contrast, IPW estimates are biased across all PS models, including LR and LDA. This pattern is consistent with theoretical results: IPW relies entirely on the correct specification of the PS, whereas AIPW retains consistency when the outcome model is correctly specified, and RSM remains unbiased when its mean structure is correctly modeled \cite{robins1994estimation, bang2005doubly}. The stability of AIPW in this setting highlights its double robustness property, whereas the failure of IPW underscores its vulnerability to propensity-score misspecification.

When the outcome model is misspecified but the PS model is correctly specified (Scenario 3), AIPW again produces unbiased estimates with comparatively low RMSE and SE. IPW performs well when the PS is estimated using parametric or semiparametric methods such as LR and LDA, but exhibits bias and increased variability when random forests or SVM are used. This reflects the fact that flexible machine learning methods may not target the true PS consistently in finite samples, even when the underlying model is correctly specified. As expected, RSM showed substantial bias in this setting due to misspecification of the outcome regression, confirming that outcome-only approaches are not robust to violations of the outcome model assumptions \cite{vanderlaan2011}.

Finally, when both the PS and outcome models are misspecified (Scenario 4), AIPW exhibits lower bias across all settings, whereas both IPW and RSM exhibit substantial bias, inflated RMSE, and wider confidence intervals. These results provide strong empirical support for the double robustness property of AIPW: consistency is preserved as long as at least one of the two nuisance models is correctly specified, whereas estimators relying on a single model fail when that model is incorrect \cite{bang2005doubly, tsiatis2006semiparametric}.


Across all four model specification regimes, increasing the sample size from $n=200$ to $n=1000$ led to marked improvements in estimator performance, with reductions in bias, RMSE, SE, and confidence interval width, as expected from asymptotic theory. Higher correlation among covariates ($\rho = 0.7$ compared to $\rho = 0.2$) consistently results in improved estimation accuracy. This likely arises because stronger correlation structures enhance the predictive strength of included covariates, thereby stabilizing both PS and outcome model estimation. These trends are clearly visible in the boxplots, where distributions become more concentrated around the true treatment effect as either sample size or correlation increases.

\section{Application to Real Data} \label{application}

We present two real-world case studies to illustrate the comparison: ACTG 175, a double-blind randomized clinical trial, and ADNI, a retrospective study of Alzheimer’s disease. In both analyses, estimated PSs are truncated to the interval [0.025, 0.975]—values below 0.025 are set to 0.025 and values above 0.975 are set to 0.975—to improve numerical stability.

\subsection{ACTG175 study}

As a first example, we analyze data from ACTG Protocol 175, a randomized, double-blind trial comparing four antiretroviral regimens in HIV-1–infected adults with CD4 counts between 200 and 500 per cubic millimeter \cite{hammer1996trial}. Because treatment was randomly assigned, confounding due to measured covariates is not expected, and PS adjustment is therefore unnecessary for identification. We therefore use this study as a methodological benchmark to assess the performance of IPW, AIPW, and RSM in a setting where all estimators should recover the same causal effect. Agreement across estimators and PS specifications serves as a calibration check, whereas discrepancies may reflect finite-sample instability (e.g., extreme weights). This contrasts with the ADNI analysis in Section \ref{sec:adni_study}, where lack of randomization makes such differences substantive.

This study focuses on two treatment arms: zidovudine only (Control arm) and zidovudine plus didanosine, zidovudine plus zalcitabine, or didanosine only (Treatment arm). After excluding subjects with missing 96-week CD4 counts, the analytic sample includes 654 participants (321 control, 333 treated). The outcome is CD4 count at 96 weeks, and treatment is defined as a binary indicator for the combination regimen. Baseline covariates used in both the PS and outcome models include baseline CD4 count, Karnofsky score, symptomatic status, prior antiretroviral exposure, age, sex, race, hemophilia status, and intravenous drug use. A linear model was fitted to test the quadratic effect of age, which was found to be insignificant and was subsequently excluded from the models. Continuous variables are standardized, so the average treatment effect is expressed in standard deviation units of the 96-week CD4 distribution, with positive values indicating benefit.

Table~\ref{tab:base.summ.actg} shows no significant baseline differences between treatment groups (all $p$-values $>$ 0.05), consistent with randomization. We estimate treatment effect using IPW, AIPW, and RSM, with PSs modeled via LR, RF, LDA, and SVM, yielding nine estimators.  Inference is based on 1,000 bootstrap samples, testing $H_0: \tau \leq 0$ versus $H_a: \tau > 0$.

\begin{table}[ht]
\centering
\caption{Characteristics of ACTG175 study by treatment group}
\begin{tabular}{lccc}
\hline
\textbf{Characteristics} & \textbf{Control arm} & \textbf{Treatment arm} & \textbf{$p$-value} \\
                            &$(n=321)$                            &$(n=333)$                 \\ \hline
Age, years (mean/sd) & 35.1 (8.7) & 35.1 (8.9) & 0.969 \\
Weight, kg (mean/sd) & 75.1 (11.9) & 73.8 (13.0) & 0.176 \\
Baseline CD4 count (mean/sd) & 365.1 (115.5) & 348.2 (126.4) & 0.074 \\
Baseline CD8 count (mean/sd) & 1009.4 (503.6) & 1022.8 (487.3) & 0.730 \\
Karnofsky score (mean/sd) & 95.8 (5.8) & 96.0 (5.5) & 0.633 \\
CD4 count at 96 weeks (mean/sd) & 287.6 (166.4) & 341.3 (173.6) & $<0.0001$ \\
Gender, Male (\%) & 81.9 & 82.9 & 0.828 \\
Race, White (\%) & 74.5 & 75.7 & 0.787 \\
Symptomatic status, Yes (\%) & 15.6 & 19.8 & 0.188 \\
Hemophilia, Yes (\%) & 7.2 & 9.3 & 0.393 \\
Homosexual activity, Yes (\%) & 65.1 & 65.8 & 0.925 \\
IV drug use, Yes (\%) & 9.0 & 13.2 & 0.116 \\
Antiretroviral, Experienced (\%) & 57.6 & 61.0 & 0.431 \\
\hline
\end{tabular}
\label{tab:base.summ.actg}
\vspace{0.5em}
\footnotesize
Note that $p$-values were calculated using Chi-squared test for categorical variables and the two-sample $t$-test for continuous variables.
\end{table}

\begin{figure} [ht]
\centering
\includegraphics[width=12cm,height=8cm]{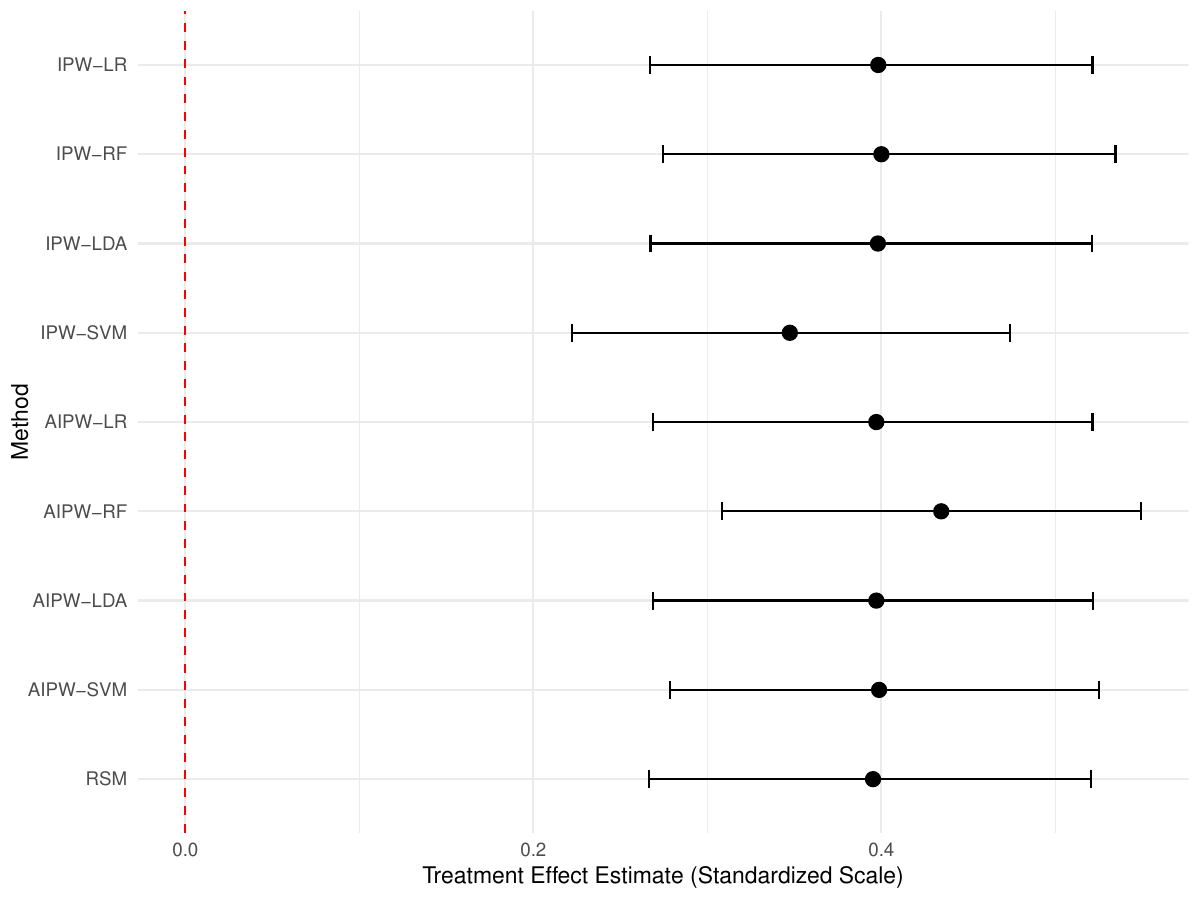}%
\caption{Estimate of treatment effect on the change of CD4 count across various methods}%
\label{fig:ACTG175}%
\end{figure}

Figure~\ref{fig:ACTG175} shows that all estimators yield positive and statistically significant effects, with 95\% confidence intervals excluding zero. The estimated treatment effect ranges from approximately 0.34 to 0.43 standard deviation units, indicating a clinically meaningful improvement in CD4 count at week 96 with combination therapy. Estimates are highly consistent across PS models within each estimator class, and AIPW and RSM results are nearly identical.

This strong agreement across various methods is expected under randomization, where all estimators target the same causal effect and differ only by sampling variability. In contrast to Section \ref{sec:adni_study}, where discrepancies reflect model dependence and limited overlap, these results provide a calibration benchmark demonstrating that the estimators behave as expected when identification assumptions hold. A slight upward deviation for AIPW with random forests is observed, but it remains within bootstrap variability and does not affect the overall conclusion.

\subsection{ADNI study} \label{sec:adni_study}
As a second example, we analyze data from a retrospective observational study that is reorganized to resemble a randomized controlled trial (RCT). Specifically, we use data from the Alzheimer’s Disease Neuroimaging Initiative (ADNI), a widely used longitudinal resource for Alzheimer’s disease research. From this database, we construct a quasi-experimental exposure–control framework. Although the resulting design does not fully replicate the conditions of a true randomized trial, it provides a suitable and informative setting for our research purpose.

The primary outcome is derived from the Alzheimer’s Disease Assessment Scale–Cognitive Subscale (ADAS13), which is commonly regarded as a benchmark outcome for evaluating the effectiveness of anti-dementia interventions \cite{kueper2018alzheimer}. The ADAS13 score aggregates across 13 cognitive tasks, with larger values indicating greater cognitive impairment. For each participant, we define the outcome as the change in ADAS13 from baseline to week 52, with positive values indicating improvement in cognitive function. Baseline covariates considered in the analysis include age, sex, Rey Auditory Verbal Learning Test (RAVLT) performance, APOE4 genotype, and Mini–Mental State Examination (MMSE) score. Although five diagnostic categories are available in the ADNI database, we focus on three clinically relevant groups—cognitively normal (CN), late mild cognitive impairment (LMCI), and Alzheimer’s disease (AD)—to define the exposure variable. Participants classified as CN comprise the control group, while those diagnosed with LMCI or AD are combined to form the exposure group. APOE4 status is categorized as 0, 1, or 2 alleles, MMSE scores range from 18 to 30, and RAVLT performance is summarized by averaging its immediate recall, learning, and forgetting components into a single composite measure.

The analysis uses data from three ADNI enrollment phases (ADNI1, ADNI2, and ADNI3), yielding a total sample of 1,241 subjects. The primary goal of this study is to estimate the exposure effect on cognitive change using multiple causal inference estimators and to assess the impact of alternative statistical and machine learning approaches for PS estimation. We first fit a standard logistic regression model including all baseline covariates. Because the quadratic term for age was statistically significant, it was retained in the model. Consequently, age and its quadratic term were included in both the PS and outcome models. All covariates and outcome variables were standardized to ensure comparability across scales prior to analysis.

\begin{table}[ht]
\centering
\caption{Characteristics of the ADNI study by exposure and control groups}
\medskip
\begin{tabular}{lccc} \hline
\textbf{Characteristics}    & \textbf{Exposure group} & \textbf{Control group} & \textbf{$p$-value}\\ 
                            &$n=842$                            &$n=399$                 \\ \hline 
Gender, Male (\%)            & 59.74 & 50.13 & 0.002\\
Age, years (mean/SD)          &  74.27 (7.61) & 74.62 (5.79) & 0.359\\ 
Baseline RAVLT (mean/SD)    & 12.12 (3.95) & 18.03 (3.54) & $<0.0001$\\ 
APOE4, $0$ (\%)              & 41.21 & 71.93 & $<0.0001$\\ 
\hspace{10.5mm}  $1$ (\%)   &  42.87 & 25.06 \\ 
Baseline MMSE  (mean/SD)    &  25.88 (2.67) & 29.08 (1.13) & $<0.0001$\\
ADAS13 score difference (mean/SD) &  -2.48 (5.97) & 0.55 (3.87) & $<0.0001$\\ 
\hline
\end{tabular}
\label{tab:base.summ}
\vspace{0.5em}
\footnotesize
Note that $p$-values were calculated using Chi-squared test for categorical variables and the two-sample $t$-test for continuous variables.
\end{table}

We apply the methods described above to estimate the magnitude of the exposure effect. The hypothesis test is formulated as $H_0: \tau = 0$ versus $H_a: \tau \neq 0$, where $\tau$ denotes the exposure effect, defined as the difference in the mean change of ADAS13 scores between the exposed and control groups. Larger positive values of $\tau$ indicate greater improvement in cognitive function. We utilize 1000 bootstrap samples to calculate the standard error of the estimates that will be used for inference. Table \ref{tab:base.summ} shows the distribution of baseline covariates between the exposed and control groups.

\begin{figure} [ht]
\centering
\includegraphics[width=12cm,height=8cm]{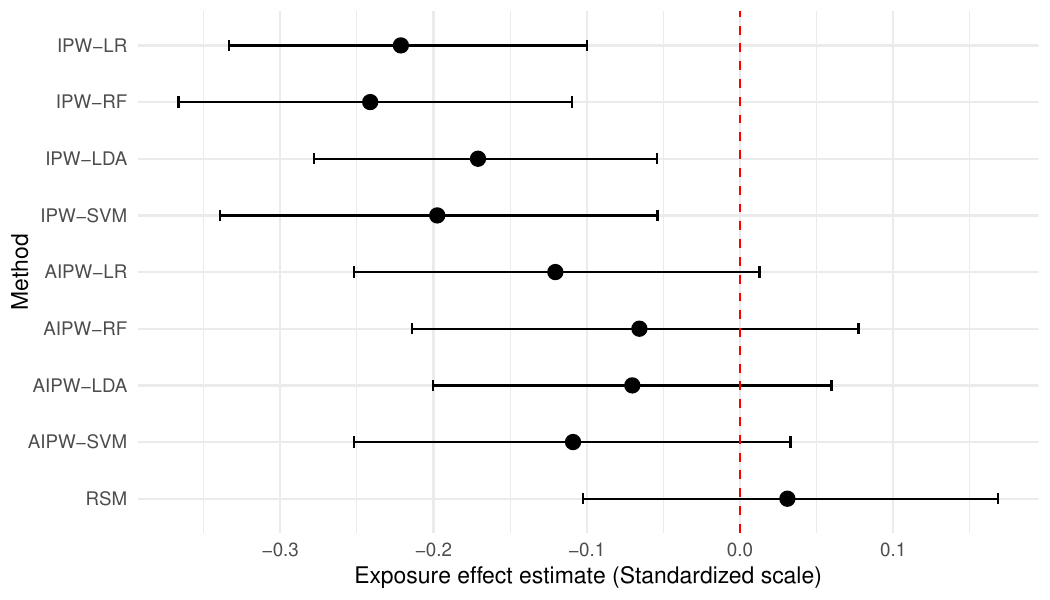}%
\caption{Estimate of exposure effect on the change of ADAS13 score across various methods}%
\label{fig:ADNI}%
\end{figure}

Figure \ref{fig:ADNI} presents estimated exposure effects on the standardized change in ADAS13 scores obtained using multiple causal estimators combined with different PS models. Across all IPW estimators, the estimated exposure effects are consistently negative and statistically significant, regardless of whether the PS model is estimated using LR, RF, LDA, or SVM. Individuals in the exposed group (LMCI/AD) experienced, on average, a greater decline or smaller improvement in cognitive function over 52 weeks compared to cognitively normal controls. The consistency of the IPW results across diverse PS models suggests that the direction of the estimated effect is robust to the choice of PS specification when weighting alone is used. In contrast, the AIPW estimators yield effect estimates that are smaller in magnitude and whose confidence intervals include zero for all PS models considered. AIPW uses outcome model information to partially correct for PS misspecification and extreme weights, yielding more attenuated and stable estimates. By leveraging outcome information, AIPW mitigates bias arising from PS misspecification and reduces sensitivity to extreme weights, leading to more conservative and stable estimates. The observed discrepancy between IPW and AIPW highlights the well-known variance–bias trade-off. While IPW may amplify residual confounding or lack of overlap through large weights, AIPW partially corrects for these issues, resulting in estimates closer to the null.

The RSM estimator yields an exposure effect close to zero with a confidence interval that includes the null, indicating no statistically significant difference in cognitive change between the exposed and control groups under this approach. Compared with IPW and AIPW, the RSM result is substantially attenuated, reflecting its reliance on outcome modeling rather than reweighting the covariate distribution. Overall, these results demonstrate that causal effect estimates in this quasi-experimental ADNI setting are sensitive to the choice of estimator, even when the PS model is varied. While different PS models yield broadly similar conclusions within each estimator class, the contrast between IPW, AIPW, and RSM underscores the importance of jointly considering PS modeling and effect estimation strategies in observational causal analyses.

\section{Discussion} \label{discussion}
Our primary goal was to investigate how different PS estimation strategies influence the performance of causal effect estimators under varying levels of model misspecification. In particular, we examined the impact of classical parametric models and flexible machine learning approaches for PS estimation when combined with IPW, AIPW, and RSM. By considering multiple simulation settings, including correct and misspecified models, varying sample sizes, and different levels of covariate correlation, this study aimed to provide practical guidance on the reliability and robustness of commonly used causal estimators in realistic data settings.

Our simulation results show clear differences in estimator performance across methods and settings. When both the PS and outcome models were correctly specified, AIPW and RSM consistently produced unbiased estimates with low variability, whereas IPW was sensitive to the choice of PS model and exhibited bias when the PS was estimated using flexible machine learning methods such as RF and SVM. Under PS misspecification, AIPW continued to perform well due to its double robustness property, while IPW and RSM suffered from increased bias. When the outcome model was misspecified, RSM failed as expected, whereas AIPW remained unbiased and IPW performed well only when the PS was estimated using correctly specified parametric models. 

Our results provide important insight into the role of ensemble learning methods, which combine the strengths of multiple models, for PS estimation. Ensemble approaches are designed to optimize predictive accuracy by aggregating information across multiple models, making them attractive for estimating treatment assignment mechanisms in complex settings. However, as noted in prior work, PS estimation requires more than accurate prediction; it also requires stable estimation of treatment probabilities away from the boundaries of zero and one \cite{kang2007demystifying,lee2010improving}. Moreover, when used within AIPW, the flexibility of ensemble learners can be exploited without compromising estimator robustness, consistent with recommendations in the targeted learning literature \cite{vanderlaan2011}.

Tree-based methods such as random forests might offer advantages in settings where the true treatment assignment mechanism is highly nonlinear or involves complex interactions. Previous studies have shown that such methods can improve nuisance function estimation in causal inference when properly integrated into doubly robust or targeted estimators \cite{athey2019generalized, chernozhukov2018double}. Our results support this perspective. Although random forests performed poorly within IPW, they yielded reliable estimates when paired with AIPW, suggesting that their strength lies in flexible nuisance modeling rather than standalone weighting.

Although our simulation studies focus on low-dimensional settings, real-world applications often involve substantially higher-dimensional data.  As the dimensionality of covariates increases, both PS and outcome model estimation become more challenging, increasing the risk of overfitting and extreme estimated probabilities. In such settings, preliminary variable selection procedures, such as LASSO \citep{tibshirani1996regression} or stability selection–based methods \citep{meinshausen2010stability, shah2013variable, das2026identifying} may be applied prior to fitting the PS model to identify the most relevant covariates and improve model stability. Incorporating such techniques into PS estimation may improve stability, particularly when combined with doubly robust estimators. 

Finally, our results highlight a broader principle in causal inference: flexibility in nuisance modeling must be balanced with robustness of the estimator. Machine learning methods are powerful tools for capturing complex data structures, but their use in causal estimation requires thoughtful integration with estimators that are resilient to model misspecification. Doubly robust methods such as AIPW provide a natural framework for this integration, offering protection against errors in either the PS or outcome model.

Although this study provides concrete guidance on selecting PS models and causal effect estimators under various settings, it has several limitations. First, the simulation settings considered moderate-dimensional covariates and additive outcome models; performance may differ in ultra–high-dimensional settings or under stronger nonlinear outcome structures. Second, our analysis assumes that all relevant confounding variables are fully observed and correctly measured. In practice, however, unmeasured or poorly measured confounders are often present, which may introduce residual bias and compromise the validity of causal effect estimates. Third, inference was based on large-sample approximations, and alternative variance estimation strategies may be needed in small samples. Fourth, we focused only on the simple nature of data, without incorporating complex or non-linear effects.

In conclusion, our results demonstrate that the choice of PS estimation method has a substantial impact on causal effect estimation, particularly for weighting-based estimators. While flexible machine learning models offer advantages in capturing complex treatment assignment mechanisms, their direct use within IPW can be problematic. Doubly robust estimators such as AIPW provide a reliable and practically appealing alternative, maintaining stable and accurate performance across a wide range of model specifications. These findings support the use of doubly robust frameworks when integrating modern machine learning tools into causal inference.

These considerations are particularly relevant for clinical trials and regulatory science, where methodological transparency and robustness are essential for credible decision-making. As the use of real-world and externally sourced data continues to expand, there is a growing need for principled approaches that can accommodate complex data structures while maintaining interpretability and stability. The framework considered in this study aligns with current regulatory emphasis on reproducibility and sensitivity to modeling assumptions, and may inform the development of analysis strategies that are both scientifically rigorous and practically implementable in regulatory settings.

\backmatter

\section*{Supplementary Information}
Additional simulation results are provided in the online supplementary material.

\section*{Acknowledgments} 
Data collection and sharing for this project was funded by the Alzheimer's Disease Neuroimaging Initiative (ADNI) (National Institutes of Health Grant U01 AG024904) and DOD ADNI (Department of Defense award number W81XWH-12-2-0012). ADNI is funded by the National Institute on Aging, the National Institute of Biomedical Imaging and Bioengineering, and through generous contributions from the following: AbbVie, Alzheimer’s Association; Alzheimer’s Drug Discovery Foundation; Araclon Biotech; BioClinica, Inc.; Biogen; Bristol-Myers Squibb Company; CereSpir, Inc.; Cogstate; Eisai Inc.; Elan Pharmaceuticals, Inc.; Eli Lilly and Company; EuroImmun; F. Hoffmann-La Roche Ltd and its affiliated company Genentech, Inc.; Fujirebio; GE Healthcare; IXICO Ltd.; Janssen Alzheimer Immunotherapy Research \& Development, LLC.; Johnson \& Johnson Pharmaceutical Research \& Development LLC.; Lumosity; Lundbeck; Merck \& Co., Inc.; Meso Scale Diagnostics, LLC.; NeuroRx Research; Neurotrack Technologies; Novartis Pharmaceuticals Corporation; Pfizer Inc.; Piramal Imaging; Servier; Takeda Pharmaceutical Company; and Transition Therapeutics. The Canadian Institutes of Health Research is providing funds to support ADNI clinical sites in Canada. Private sector contributions are facilitated by the Foundation for the National Institutes of Health (\url{www.fnih.org}). The grantee organization is the Northern California Institute for Research and Education, and the study is coordinated by the Alzheimer’s Therapeutic Research Institute at the University of Southern California. ADNI data are disseminated by the Laboratory for Neuro Imaging at the University of Southern California.

\section*{Declarations}
\begin{itemize}
\item Funding: Not applicable.
\item Conflict of interest/Competing interests: Authors declare no potential conflict of interest.
\item Ethics approval and consent to participate: Not applicable.
\item Data availability: ACTG175 data can be obtained from this study \cite{hammer1996trial} and \href{https://adni.loni.usc.edu/}{ADNI} data can be obtained upon request.
\item Code availability: All code is written in statistical software R 4.5 and the code
can be downloaded from the \href{https://github.com/apustat/Propensity-score-model-comparison-under-model-misspecification/}{GitHub}.
\item Declaration of Generative AI and AI-assisted technologies:
During the preparation of this manuscript, the authors used ChatGPT-5 in order to proofread and improve the language and readability. After using this tool/service, the authors reviewed and edited the content as needed and take full responsibility for the content of the published article.
\end{itemize}

\begin{appendices}

\end{appendices}
\bibliography{sn-bibliography}

\end{document}